\documentstyle[preprint,floats,aps,epsf,prb]{revtex}

\newcommand \be  {\begin{equation}}
\newcommand \ee  {\end{equation}}
\newcommand \bea {\begin{eqnarray} }
\newcommand \eea {\end{eqnarray}}
 
\begin{document}
\draft
\title{Ward identities for disordered metals and superconductors} 
\author{Revaz Ramazashvili}
\address{Materials Science Division,  Argonne National Laboratory, Argonne, IL 60439,
USA.} 
\maketitle
\date{\today}
\maketitle
\begin{abstract}
This article revisits Ward identities for disordered interacting 
normal metals and superconductors. It offers a simple derivation 
based on gauge invariance and recasts the identities in a new form 
that allows easy analysis of the quasiparticle charge conservation 
(as e.g. in a normal metal) or non-conservation (as e.g. in a d-wave 
superconductor). 
\end{abstract}
\vskip 0.2 truein

\section{Introduction}

Interplay of interaction and disorder remains one of the central topics 
in condensed matter physics. Given the complexity of the problem, constraints 
imposed by symmetries acquire particular importance. An example of such 
a constraint is given by Ward identities. In the early days of many-body 
theory, Ward identities were used to establish key properties of the Fermi 
liquid theory. \cite{agd} In the context of the CPA approximation for a 
disordered non-interacting metal, similar identities were derived by Velick\'{y}. 
\cite{velicky} In the theory of superconductivity, Ward identities were used 
early on to establish gauge invariance of the electromagnetic response. \cite{s} 
Subsequently, they were employed by D. Vollhardt and P. W\"{o}lfle \cite{VW1} 
in a self-consistent theory of the Anderson transition, by F. Wegner, \cite{W1} 
A. J. McKane and M. Stone \cite{MKS1} in the sigma model approach to localization, 
and by C. Castellani \emph{et al.} \cite{C1} in an early treatment of an 
interacting disordered metal. Very recently, T. R. Kirkpatrick and D. Belitz 
\cite{KB1} invoked the Ward identities in an attempt to resolve the issue 
of decoherence at zero temperature. 

This paper revisits the Ward identities for disordered interacting normal 
metals and superconductors. Using gauge invariance, it derives the identities 
in a new form that makes quasiparticle charge conservation (as e.g. in a 
normal metal) or absence thereof (as e.g. in a d-wave superconductor) explicit. 
In a normal metal, the identity takes a particularly simple form: 

\[
\Lambda _{RA}(\omega ,\omega ';p,p) =
 -\frac{2i\Sigma _{R}''(\omega ,p)}{\omega -\omega '},
\]
where \( \Lambda _{RA} \) is the disorder average of the retarded-advanced
charge density vertex correction at zero momentum transfer \( Q=p-p=0 \) and
small frequency transfer \( \Omega =\omega -\omega '\ll \omega ,\omega ' \),
and \( \Sigma _{R}''(\omega ,p) \) is the imaginary part of the retarded quasiparticle
self energy, which is proportional to the quasiparticle scattering rate. The
vertex \( \Lambda _{RA} \) is closely related to the correlation function of
the quasiparticle charge density, and the \( 1/(\omega -\omega ') \) behavior
of the vertex at low frequency transfer and zero momentum transfer points to
quasiparticle charge conservation and its diffusive propagation. By contrast,
in a d-wave superconductor, the Ward identity reflects the fact that impurity
scattering causes exchange of charge between the quasiparticle subsystem and
the condensate, which leads to non-conservation of the quasiparticle charge. 

The structure of the paper is as follows. Section II gives a detailed derivation
of the Ward identities for a disordered interacting normal metal. 
Section III briefly discusses the Ward identities for an s-wave superconductor 
in the approximation of a spatially uniform gap. Section IV derives the Ward 
identifies for a disordered d-wave superconductor in the same approximation, 
and Section V illustrates the meaning of the identity by explaining how, 
in a d-wave superconductor, the impurity scattering leads to exchange of charge 
between the quasiparticle subsystem and the condensate. 
Section VI presents a summary and a brief discussion of the results.

\section{Ward identities for a normal metal}

Consider a disordered interacting normal metal with the Matsubara action

\begin{eqnarray*}
S=\int d\tau \int dr\psi ^{+}(r,\tau )\left[ i\hbar \partial _{\tau }-\xi (-i\hbar \vec{\nabla }-\frac{e}{c}\vec{A})+e\phi (r,\tau )-u(r)\right] \psi (r,\tau )- &  & \\
-\int d\tau dr\int d\tau 'dr'\psi _{\alpha }^{+}(r,\tau )\psi _{\beta }(r,\tau )V_{\alpha \beta \gamma \delta }(\tau -\tau ',r-r')\psi _{\gamma }^{+}(r',\tau ')\psi _{\delta }(r',\tau '), &  & 
\end{eqnarray*}
 where \( \psi ^{+} \) (\( \psi  \)) are the electron creation (annihilation)
operators, \( \vec{A} \) is the electromagnetic vector potential, \( \phi (r,\tau ) \)
is the scalar potential and \( u(r) \) is the impurity potential. This action
respects the continuous gauge symmetry

\[
\psi \rightarrow e^{i\chi (r,\tau )}\psi ;\; \; \vec{A}\rightarrow \vec{A}+\frac{\hbar c}{e}\vec{\nabla }\chi ;\; \; \phi \rightarrow \phi +\frac{\hbar }{e}\partial _{\tau }\chi ,\]
of which the sought Ward identities are a consequence. To establish the scheme
used throughout the rest of this article, I present below a detailed derivation. 

Everywhere hereafter, only infinitesimal time dependent spatially uniform transformations
\( \psi _{\alpha }(r,\tau )\rightarrow e^{i\chi (\tau )}\psi _{\alpha }(r,\tau ) \)
will be considered. Under such a transformation, the Green function changes
according to 

\[
G_{\alpha \beta }(r,r',\tau -\tau ')\rightarrow e^{i\chi (\tau )}G_{\alpha \beta }(r,r',\tau -\tau ')e^{-i\chi (\tau ')}\]
 and thus, to first order in \( \chi  \), its variation equals
\[
\delta G_{\alpha \beta }(r,r',\tau -\tau ')\approx i[\chi (\tau )-\chi (\tau ')]G_{\alpha \beta }(r,r',\tau -\tau ').\]
 On the other hand, the same transformation induces extra terms in the action
due to the presence of the temporal derivative. Hence, the very same variation
of the Green function can also be calculated by perturbation theory. The crucial
point is that the four-fermion interaction term in the action is invariant under
the gauge transformation and, therefore, does not contribute to the perturbative
correction to the Green function. Thus, to first order in infinitesimal \( \chi  \),
the same correction to \( G \) is equal to

\[
\delta G_{\alpha \beta }(x,x',\tau -\tau ')=-i\int dtdr\langle \psi _{\alpha }(x,\tau )\psi _{\gamma }^{+}(r,t)\psi _{\gamma }(r,t)\psi _{\beta }^{+}(x',\tau ')\rangle \partial _{t}\chi (t).\]

Equating the two expressions leads to the identity

\[
[\chi (\tau )-\chi (\tau ')]G_{\alpha \beta }(x,x',\tau -\tau ')=-\int dtdr\langle \psi _{\alpha }(x,\tau )\psi _{\gamma }^{+}(r,t)\psi _{\gamma }(r,t)\psi _{\beta }^{+}(x',\tau ')\rangle \partial _{t}\chi (t)\]
 for a given disorder configuration. Disorder averaging replaces the exact Green
function on the left hand side by its translationally invariant average. The
average on the right hand side can be presented as the product of the two average
Green functions plus the vertex correction term and, for \( \chi =\chi _{0}e^{i\Omega \tau } \)
with \( \Omega \rightarrow 0 \), the Fourier transformed identity takes the
form 

\[
G(i\omega +i\Omega ,p)-G(i\omega ,p)=i\Omega G(i\omega +i\Omega ,p)\left[ 1+\Lambda (i\omega ,i\omega +i\Omega ;p,p)\right] G(i\omega ,p),\]
 where \( \Lambda (i\omega ,i\omega +i\Omega ;p,p) \) is the disorder average
of the scalar vertex correction. At this point, two different types of identities
can be derived: one for the retarded-advanced vertex correction \( \Lambda _{RA}(\omega ,\omega +\Omega ;p,p) \),
and another one for the retarded-retarded vertex correction \( \Lambda _{RR}(\omega ,\omega +\Omega ;p,p) \).

\subsection{The identity for the retarded-advanced (RA) vertex}

To obtain the identities for the retarded-advanced vertex, choose \( i\omega  \)
to be in the lower half-plane and \( i\omega +i\Omega  \) in the upper half-plane.
Then, upon analytic continuation \( i\omega \rightarrow \omega \pm i0 \),
\( G(i\omega ) \) transforms into \( G_{A}(\omega -i0) \), whereas \( G(i\omega +i\Omega ) \)
transforms into \( G_{R}(\omega +\Omega +i0) \). The identity then takes the
form

\[
G_{R}^{-1}(\omega +\Omega +i0,p)-G_{A}^{-1}(\omega -i0,p)=\Omega [1+\Lambda _{RA}(\omega +\Omega ,\omega ;p,p)].\]
 The disorder averaged Green function reads \( G_{A/R}^{-1}(\omega ,p)=\omega -\Sigma _{A/R}(\omega ,p)-\xi (p), \)
where \( \Sigma _{A/R}(\omega ,p) \) is the advanced/retarded self energy.
Using the relation \( \Sigma _{R}(\omega ,p)=\Sigma ^{*}_{A}(\omega ,p) \)
and assuming that the derivative \( \partial _{\omega }\Sigma _{R/A}(\omega ,p) \)
is non-singular, for small \( \Omega =\omega -\omega '\rightarrow 0 \) one
finds 
\begin{equation}
\label{RA-identity}
-2i\Sigma _{R}''(\omega ,p)=[\omega -\omega ']\Lambda _{RA}(\omega ,\omega ';p,p).
\end{equation}
Identifying \( 2\Sigma _{R}''(\omega ) \) with the scattering rate \( 1/\tau  \),
one immediately recognizes in \( \Lambda _{RA} \) the zero momentum transfer
(\( Q=0 \)) limit of the charge density vertex \( D(\omega -\omega ',Q) \)
\cite{sa} 

\[
D(\omega -\omega ',Q)=\frac{1}{i(\omega -\omega ')\tau +DQ^{2}\tau }.\]
where \( D \) is the diffusion coefficient. For a non-interacting disordered
metal, \( D(\omega -\omega ',Q) \) is commonly obtained by a direct calculation,
\cite{sa} first finding self-consistently the impurity self energy, and then
summing the ladder series for the vertex. In the presence of interactions, diagrammatic
treatment becomes much more involved, while the present derivation appeals only
to gauge invariance and is insensitive to turning on the interaction.

\subsection{The identity for the retarded-retarded (RR) vertex}

By contrast with the identity just derived, the identity for the retarded-retarded
vertex can be found in textbooks, and I present its derivation here only for
completeness. In this case, it is convenient to choose both \( i\omega  \)
and \( i\omega +i\Omega  \) in the same (say, the upper) half-plane. Upon analytic
continuation and multiplication by \( G^{-1}(i\omega ) \) and \( G^{-1}(i\omega +i\Omega ) \),
the identity takes the form

\[
G_{R}^{-1}(\omega +\Omega +i0,p)-G_{R}^{-1}(\omega +i0,p)=\Omega [1+\Lambda _{RR}(\omega +\Omega ,\omega ;p,p)],\]
which, to first order in \( \Omega \rightarrow 0 \), leads to the standard
relation between the energy derivative of the retarded self energy and the retarded-retarded
vertex \cite{Mahan} : 

\begin{equation}
\label{RR-identity}
\partial _{\omega }\Sigma _{R}(\omega )=\Lambda _{RR}(\omega ,\omega ;p,p).
\end{equation}

\section{Ward identity for an s-wave superconductor}

In the Nambu notations, the BCS Hamiltonian of an s-wave superconductor reads 

\[
H=\int dr\Psi ^{\dagger }\left[ \tau _{3}\xi (\vec{p}-\frac{e}{c}\vec{A}\tau _{3})+\tau _{1}\Delta (r)+\tau _{3}e\phi +\tau _{3}u\right] \Psi .\]
 Here \( \Psi ^{\dagger }\equiv (\psi _{\uparrow }^{\dagger },\psi _{\downarrow }) \)
is the Nambu spinor, \( \tau _{i} \) are the Nambu matrices, and \( r \) denotes
the center of mass coordinate of a Cooper pair. The pair field \( \Delta (r) \)
has been chosen real for the sake of simplicity. 

A gauge transformation takes the form \( \Psi \rightarrow \exp \left[ i\tau _{3}\frac{e}{\hbar c}\chi \right] \Psi  \)
and, in addition to the standard change of potentials \( \vec{A} \) and \( \phi  \),
has to be accompanied by the pair field transformation \( \Delta \rightarrow \Delta \exp \left[ 2i\frac{e}{\hbar c}\chi \right]  \).
One then proceeds the same way as for a normal metal, with two important points
to note. The first point amounts to the approximation of a spatially uniform
gap which, along with the frequency, gets renormalized by disorder. The second
point stems from the Nambu matrix structure of the theory: the vertices, that
appear after disorder averaging of the perturbative expression for the Green
function variation, are defined in the Nambu space and thus carry a Nambu index.
For instance, the disorder averaged term arising from the temporal derivative
of \( \chi =\chi _{0}\exp [i\Omega \tau ] \) has the form  

\[
\int dy\langle \psi (x)\bar{\psi }(y)i\Omega \tau _{3}\psi (y)\bar{\psi }(x')\rangle \rightarrow \int dydzdz'G(x,z)i\Omega \left[ \tau _{3}\delta (z-y)\delta (y-z')+\langle \tau _{3}\rangle (z,y,z')\right] G(z',x'),\]
 where the vertex correction \( \langle \tau _{3}\rangle (z,y,z') \) appears
as a result of disorder dressing of the corresponding bare vertex, the latter
being simply the Nambu matrix \( \tau _{3} \). The disorder averaged Green
function has the form \cite{agd}

\[
G(p,\omega )=[i\tilde{\omega }\tau _{0}-\xi (p)\tau _{3}-\tilde{\Delta }\tau _{1}]^{-1},\]
 where \( \tilde{\omega } \) and \( \tilde{\Delta } \) are the renormalized
frequency and the gap amplitude, which yields the Ward identity 

\[
\left[ i\tilde{\omega }-i\tilde{\omega }'\right] \tau _{3}-i\left[ \tilde{\Delta }_{\omega }+\tilde{\Delta }_{\omega '}\right] \tau _{2}=[i\omega -i\omega '][\tau _{3}+\langle \tau _{3}\rangle ]-2i\Delta [\tau _{2}+\langle \tau _{2}\rangle ].\]
 This, in turn, indicates a diffusion pole in the quasiparticle charge density
vertex correction \( \langle \tau _{3}\rangle _{RA} \) :

\[
\frac{-2i\Sigma ''_{R}(\omega ,p)}{\omega -\omega '}\tau _{3}=\langle \tau _{3}\rangle _{RA},\]
where \( \Sigma _{R}''(\omega ,p) \) is the imaginary part of the retarded
self energy renormalization of the frequency, the notation is chosen to coincide
with the normal metal limit.

\section{Ward identity for a d-wave superconductor:}

In a d-wave superconductor, the situation turns out to be quite different. The
BCS Hamiltonian of a d-wave superconductor reads 

\[
H=\int dr\Psi ^{\dagger }\left[ \tau _{3}\xi (\vec{p}-\frac{e}{c}\vec{A}\tau _{3})+\tau _{3}e\phi +\tau _{3}u\right] \Psi +\int dRdr\Psi ^{\dagger }(R+\frac{r}{2})\tau _{1}\Delta (R,r)\Psi (R-\frac{r}{2}),\]
where the pair field \( \Delta (R,r) \) has been chosen real and having d-wave
angular dependence on the relative coordinate \( r \), and \( R \) denotes
the center of mass coordinate of a Cooper pair. 

As in the s-wave case, the Hamiltonian respects the gauge symmetry, and the
identities can be obtained similarly, with one important difference: because
of the d-wave symmetry of the gap and its oscillating angular dependence, the
gap amplitude \( \Delta _{p} \), although suppressed by impurities, does not
acquire a frequency dependent renormalization. Hence the disorder average of
the quasiparticle Green function is 

\[
G(i\omega ,p)=\left[ i\tilde{\omega }-\tau _{1}\Delta _{p}-\tau _{3}\xi _{p}\right] ^{-1}.\]
Another important point is that the angular dependence of the gap leads to the
appearance of a vertex correction \( \langle \Delta _{p}\tau _{2}\rangle  \)
on the right hand side of the Ward identity, which assumes the form

\begin{equation}
\label{eq:WI-charge}
-2i\tau _{3}\Sigma ''_{R}(\omega ,p)=[\omega -\omega ']\langle \tau _{3}\rangle _{RA}+2i\langle \Delta _{p}\tau _{2}\rangle _{RA}.
\end{equation}
 As for an s-wave superconductor, \( \Sigma _{R}''(\omega ,p) \) is the retarded
self-energy renormalization of the frequency: \( \tilde{\omega }=\omega -\Sigma  \).
Due to the d-wave symmetry of the gap and its oscillatory angular dependence,
\( \langle \Delta _{p}\tau _{2}\rangle _{RA}\propto \langle \tau _{3}\rangle _{RA} \),
which leads one to conclude that the vertex correction \( \langle \tau _{3}\rangle _{RA} \)
has to remain finite as \( \omega -\omega '\rightarrow 0 \). Hence, in a disordered
d-wave superconductor, the quasiparticle charge is not conserved. Note that,
upon transition to the normal state, the quasiparticle charge diffusion mode re-appears, 
as can be seen seen by sending \( \Delta _{p} \) to zero in Eq. (\ref{eq:WI-charge})
and identifying \( \langle \tau _{3}\rangle _{RA} \) with \( \Lambda _{RA}(\omega ,\omega ';p,p) \)
of Section II.

\section{Qualitative argument for a d-wave superconductor}

The absence of quasiparticle charge conservation in a d-wave superconductor
can be understood based on a simple argument going back to the studies of charge
imbalance relaxation in superconductors. \cite{cimb} I reproduce the argument
here for the sake of completeness. Consider the Bogolyubov quasiparticle creation
operator: 

\[
\gamma ^{+}_{p\uparrow }=u_{p}c^{+}_{p\uparrow }+v_{p}c_{-p\downarrow }\; ,\; u^{2}_{p}=\frac{1}{2}[1+\frac{\xi _{p}}{\sqrt{\xi ^{2}_{p}+\Delta ^{2}_{p}}}]\; ,\; u^{2}_{p}+v^{2}_{p}=1.\]
 Impurity scattering is elastic, i.e. it conserves the quasiparticle energy
\( E_{p}=\sqrt{\xi ^{2}_{p}+\Delta ^{2}_{p}} \) . In an s-wave superconductor
with uniform gap, \( \Delta _{p} \) is a constant and, in the absence of the
Andreev scattering that turns \( \xi _{p} \) into \( -\xi _{p} \), the energy
conservation implies conservation of \( u_{p} \) and \( v_{p} \) . Hence the
impurity scattering conserves the particle-hole content of a quasiparticle,
and this leads to the effective charge conservation -- even though a Bogolyubov
quasiparticle, being a superposition of a particle and a hole, does not have
a well defined charge quantum number. The same conclusion can be reached by
considering directly the expectation value of the quasiparticle charge \( Q_{p} \)
:

\[
Q_{p}=u^{2}_{p}(+1)+v^{2}_{p}(-1)=\frac{\xi _{p}}{\sqrt{\xi ^{2}_{p}+\Delta ^{2}_{p}}}.\]
In an isotropic s-wave superconductor, the gap does not vary around the Fermi
surface and hence, in the absence of the Andreev processes, \( Q_{p} \) is
conserved by the impurity scattering, which leads to the charge diffusion pole. 

By contrast, in a d-wave superconductor, the gap \( \Delta _{p} \) is strongly
anisotropic. Thus, even in the absence of the Andreev scattering processes,
neither \( Q_{p} \) nor the moduli of the Bogolyubov factors \( u_{p} \) and
\( v_{p} \) are conserved: impurity scattering changes the particle-hole content
of a quasiparticle. Physically, this means that the impurity scattering induces
exchange of charge between the quasiparticle subsystem and the condensate. 

Indeed, this quasiparticle charge non-conservation is not a consequence of the
d-wave symmetry of the gap, but rather of the gap anisotropy around the Fermi
surface, and is present not only in other superconductors with non-trivial symmetry,
but even in s-wave superconductors with anisotropic gap. However, in the latter
case, the effect is small in the measure of the relative gap anisotropy, which
is itself reduced by disorder. As a result, the quasiparticle charge non-conservation
appears only at time scales that are long compared with the scattering time.
By contrast, in a d-wave superconductor, the gap anisotropy is large, and the
quasiparticle charge changes at the time scale of order the impurity scattering
time, which eliminates quasiparticle charge conservation at any time scale beyond
the elastic scattering time.

\section{Summary and discussion}

In this article, I revisited the Ward identities for superconductors and disordered
interacting normal metals, and presented a simple derivation based solely on
gauge invariance. The identities were recast in a new form that made quasiparticle
charge conservation (as in a normal metal or an isotropic s-wave superconductor)
or absence thereof (as in a d-wave superconductor) explicit. Using the Ward
identities, I showed how, in a d-wave superconductor, impurity scattering causes
exchange of charge between the quasiparticle subsystem and the condensate, thus
leading to the quasiparticle charge non-conservation. 


Transparency of the Ward identities is particularly appealing in comparison 
with microscopic approaches. The simplicity of the identities is insensitive 
to the strength of the impurity potential or to whether disorder has to be 
treated in the Born or in the unitary limit -- or to the presence of interaction. 
By contrast, to achieve a controllable approximation even in the Born limit, 
microscopic calculations, e.g. for a d-wave superconductor, have to resort 
to rather complex methods and/or unrealistic approximations,
such as expansion in the inverse number of gap nodes. \cite{at1} 

It is my pleasure to thank  A. J. Leggett, M. R. Norman and especially 
Yu. M. Galperin for helpful discussions. Work supported by the U. S. 
Department of Energy, Division of Basic Energy Science-Material Science 
under contract No. W-31-109-ENG-38.

\end{document}